\documentclass[ twocolumn]{aastex631}
\usepackage{amsmath, amsfonts}
\usepackage{subfigure}

\usepackage{acronym,units}
\usepackage{dcolumn}
\usepackage{bm}
\usepackage{hyperref}
\usepackage[mathlines]{lineno}
\usepackage{appendix}
\usepackage{graphicx}
\usepackage{xcolor}
\usepackage[normalem]{ulem}

\newcommand{\SPA}{School of Physics and Astronomy, Monash University, Clayton VIC 3800, Australia}
\newcommand{\OzGravMonash}{OzGrav: The ARC Centre of Excellence for Gravitational Wave Discovery, Clayton VIC 3800, Australia}

\begin{document}
\preprint{APS/123-QED}

\title{GW231123: extreme spins or  microglitches?}

\author[0000-0002-7322-4748]{Anarya Ray}
\affiliation{Center for Interdisciplinary Exploration and Research in Astrophysics (CIERA), Northwestern University, 1800 Sherman Ave, Evanston, IL 60201, USA}
\affiliation{NSF-Simons AI Institute for the Sky (SkAI), 172 E. Chestnut Street, Chicago, IL 60611, USA}

\author[0000-0001-7852-7484]{Sharan Banagiri}
\affiliation{\SPA}
\affiliation{\OzGravMonash}

\author[0000-0002-4418-3895]{Eric Thrane}
\affiliation{\SPA}
\affiliation{\OzGravMonash}

\author[0000-0003-3763-1386]{Paul D. Lasky}
\affiliation{\SPA}
\affiliation{\OzGravMonash}

\date{\today}

\begin{abstract}
The recently reported binary black hole merger, GW231123, has unusual properties that make it hard to explain astrophysically. Parameter estimation studies are consistent with maximally spinning black holes and the dimensionless spin of the more massive component is constrained to be $\chi_1\gtrsim 0.8$. Analysis of data also revealed potential systematics that could not be fully replicated with simulated studies. We explore the possibility that these measurements are biased due to unmodeled non-Gaussian noise in the detectors, and that the actual black hole spins are more modest. We present evidence for a population of \textit{microglitches} in LIGO gravitational-wave strain data that can lead to biases in the parameter estimation of short-duration signals such as GW231123. Using simulated data of a massive event like GW231123, we demonstrate how microglitches can bias our measurements of black hole spins toward $\chi\approx1$ with negligible posterior support for the true value of $\chi\approx0.7$. We develop a noise model to account for microglitches and show that this model successfully reduces biases in the recovery of signal parameters. We characterize the microglitch population in real interferometer data surrounding GW231123 and find a single detector glitch duty cycle of $0.57_{-0.19}^{+0.21}$, which implies nearly a $100\%$ probability that at least one event through the fourth gravitational wave transient catalog coincides with microglitches in two detectors. We argue that further investigations are required before we can have a confident picture of the astrophysical properties of GW231123. 

\end{abstract}


\acrodef{GW}[GW]{gravitational wave}
\acrodef{BBH}[BBH]{binary black hole}
\acrodef{LVK}[LVK]{LIGO-Virgo-KAGRA}
\section{Introduction}

On 23 November 2023, the \ac{LVK} detector network~\citep{LIGOScientific:2014pky, VIRGO:2014yos, KAGRA:2020tym} recorded \acp{GW} from GW23112\_135430 (henceforth, GW231123), the most massive \ac{BBH} merger ever detected~\citep{LIGOScientific:2025rsn}. The merger was detected by both LIGO Hanford (H1) and Livingston (L1) detectors. This event is remarkable both for the unusually large mass ($190-260 M_\odot$, 90\% credibility) and for the extreme spins. Using the \textsc{NRSur7dq4} numerical relativity surrogate waveform (abbreviated here as \textsc{nrsur})~\citep{Varma:2019csw}, they find $\chi_1 = 0.89^{+0.11}_{-0.20}$ and $\chi_2 = 0.91^{+0.09}_{-0.20}$, respectively.

 The event is difficult to analyze for a number of reasons. 
 First, the extreme properties of GW231123 presents challenges with waveform calibration, potentially introducing significant systematic errors. No waveform models are calibrated for spins $\chi > 0.8$. Mismatch tests, comparing waveform approximants to numerical relativity, suggest that \textsc{nrsur} provides a better representation of numerical relativity than other waveforms that are currently available (see Sec. 4.2 of \cite{LIGOScientific:2025rsn}). However, worryingly, the \ac{LVK} analysis of the data finds that all waveforms except \textsc{imrphenomXPHM} have a higher Bayes factor compared to \textsc{nrsur}. Moreover, the waveforms exhibit significant systematics in the recovery of the masses of the binary, to a degree that could not be replicated with simulated signals (see Appendices A and B of \cite{LIGOScientific:2025rsn}). While clear glitches are also present in the analysis window, they are not coincident with the signal and \cite{LIGOScientific:2025rsn} therefore conclude that they do not impact estimation of event properties. 

Understanding the properties of such heavy black holes is of great importance, as the black holes in GW231123 might represent a missing link between stellar-mass black holes and the more massive black holes hypothesized to exist at the centers of dense stellar clusters and which have been observed in the centers of ~\citep{Tagawa:2016mwr, Inayoshi:2019fun, Sicilia:2021gtu, Schiebelbein-Zwack:2024roj}. Hierarchical mergers in dense star clusters provides a natural explanation for forming massive, rapidly spinning black holes (see \cite{Gerosa:2021mno} for a review). However, explaining the spins of GW231123 is challenging even assuming two second generation mergers~\citep{Passenger:2025, Paiella:2025qld, Stegmann:2025cja} Alternate formation mechanisms have been suggested, including accretion-induced spin up in AGN disks, chemically homogeneous evolution, direct collapse of massive stars and stellar collisions in dense clusters~\citep{Kiroglu:2025vqy, Baumgarte:2025syh, Gottlieb:2025ugy, Croon:2025gol, DeLuca:2025fln, Bartos:2025pkv}

In this paper, we consider the hypothesis that data quality issues might be driving some of the extreme properties~\citep[see, for example, ][for more details]{Davis:2022ird, LIGO:2021ppb, LIGOScientific:2016gtq} of this event, particularly the component spins. We focus in particular on \textit{microglitches}: weak but frequent non-stationary transient noise artifacts. It has been demonstrated that measurements of black hole spins can be particularly sensitive to non-Gaussianities in the data~\citep{Udall:2025bts, Legin:2024gid, Malz:2025xdg, Raymond:2024xzj, Plunkett:2022zmx, Ghonge:2023ksb, Payne:2022spz, Hourihane:2022doe}. Our work builds on investigations in \cite{shun_memory}, which found that LIGO data can suffer from weak, non-stationary noise events, similar in morphology to a Heaviside step function. We find evidence for existence of microglitches in the data around GW231123 and demonstrate that microglitches can cause a binary with less unusual, but still astrophysically significant spins ($\chi \approx 0.7$) to exhibit apparently extremal spins similar to those observed for GW231123.

\section{Microglitch model}
\begin{figure*}
    \centering
    \includegraphics[width=0.32\linewidth]{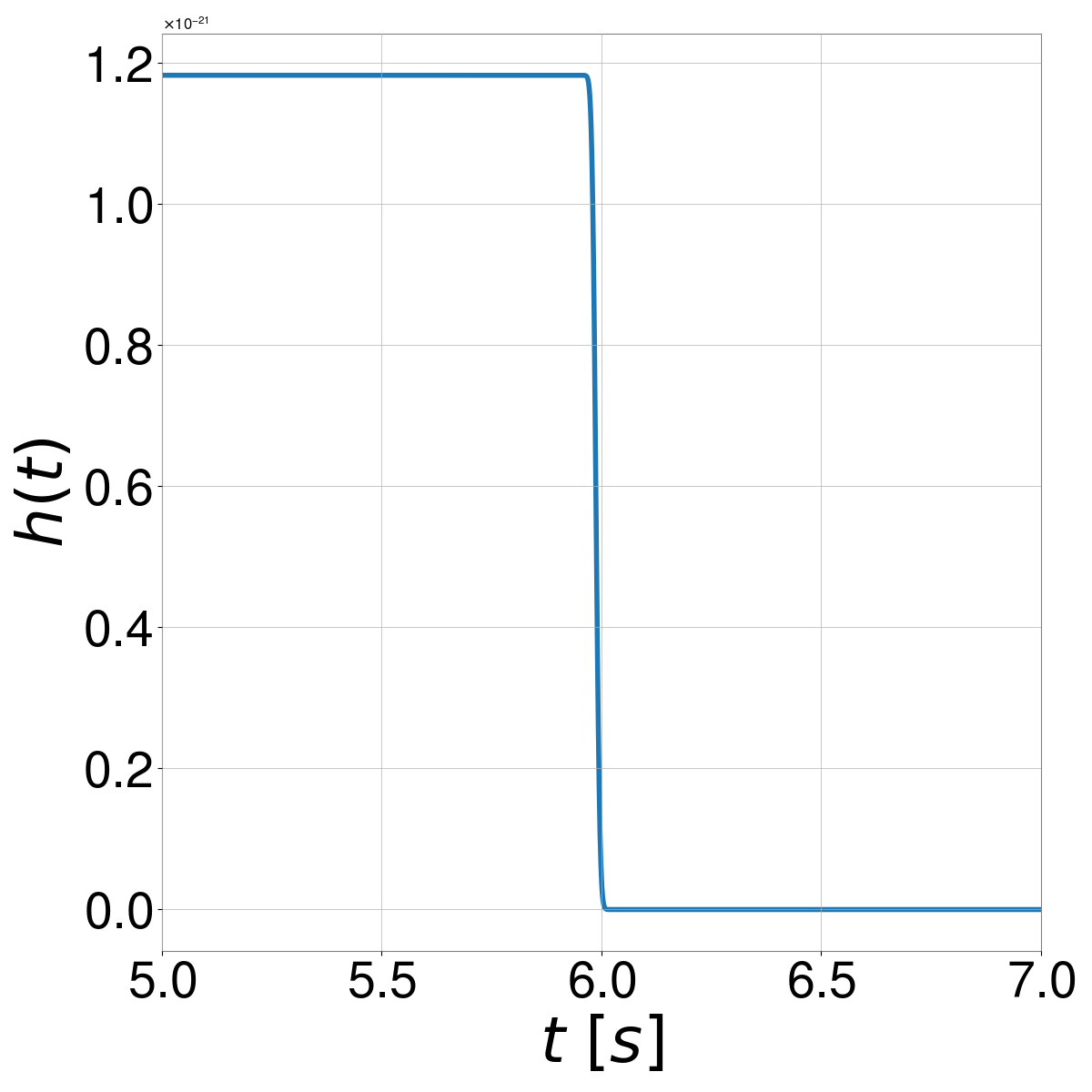}
    \includegraphics[width=0.32\linewidth]{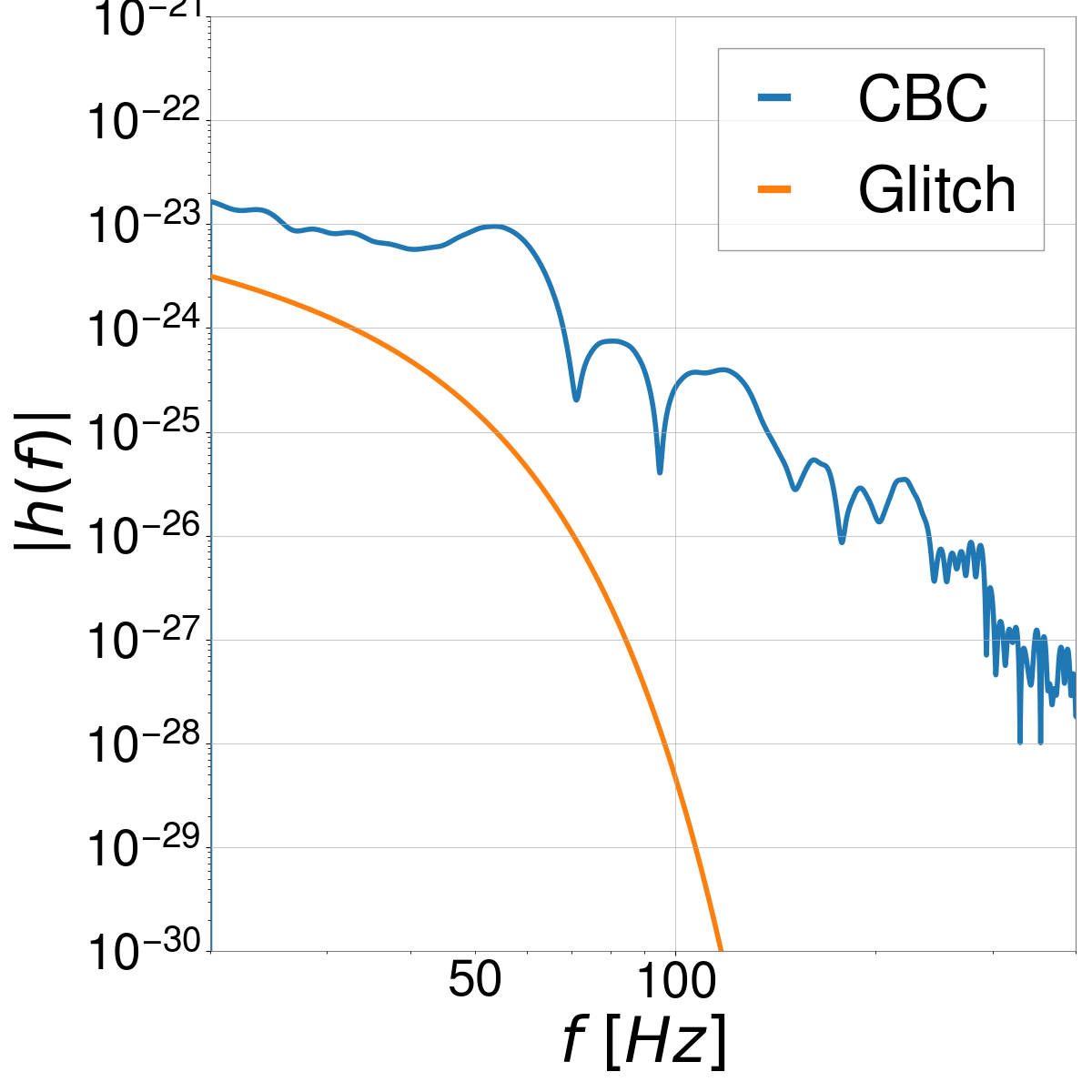}
    \includegraphics[width=0.32\linewidth]{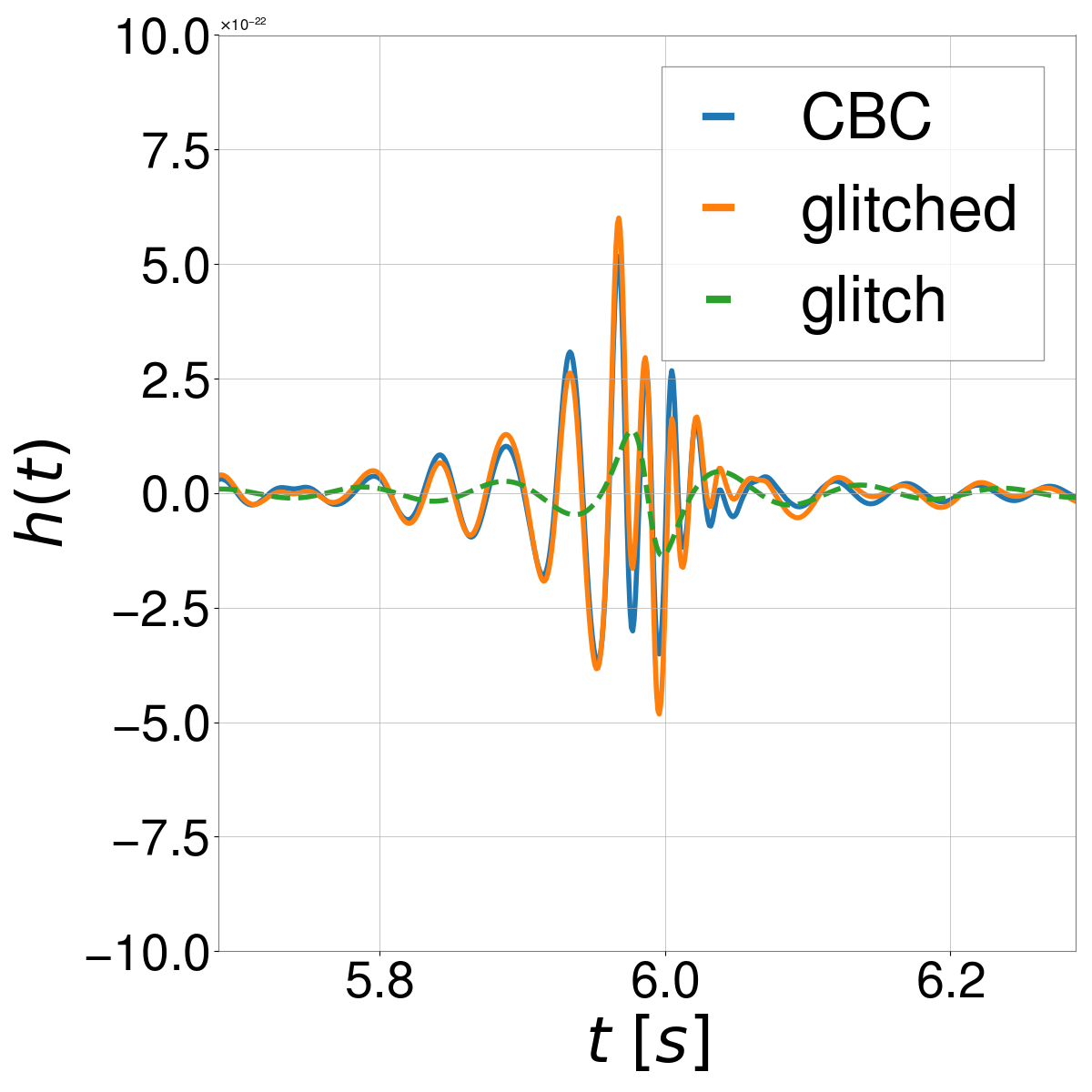}
    \caption{Schematic of our glitch model. The left panel shows the time series of an example microglitch, as defined in Eq. (1) with parameters $A = 5.8\times10^{-22}$, $\sigma=10^{-2}$, and $t_0 = 5.9s$. The middle panel shows a frequency-domain representation of the same glitch (orange) and a GW231123-like BBH signal (blue). The right panel shows the time series reconstruction of the frequency-domain glitch (green), the BBH (blue), and their sum (orange) after applying a bandpass filter (unlike the left panel which is not bandpassed). 
    }
    \label{fig:glitch_model}
\end{figure*}

Our glitch model is motivated by the \cite{shun_memory}. 
The authors of that work fortuitously discovered a population of nonstationary noise while searching for gravitational-wave memory signals, which look similar to Heaviside step functions. When band-passed to the frequencies at which LVK detectors are sensitive, the step function manifests in the time domain as a short, rapidly decaying pulse with only a few cycles; see Fig.~4 from \cite{shun_memory}.
Our phenomenological time-series model captures the salient feature of these memory bursts via the function
\begin{equation}
    g(t; A, t_0, \sigma) = \frac{A}{2} \left [1 + \text{erf} \left(- \frac{(t - t_0)}{\sigma} \right) \right] .
    \label{eq:glitch-model} 
\end{equation}
Here, $g$ is a strain time series, $A$ is the glitch amplitude, $t_0$ is the central glitch time, $\sigma$ is the decay scale of the glitch and $\text{erf}$ is the standard error function. 
In the frequency domain, 
\begin{equation}
    \tilde{g}(f; A, t_0, \sigma) = \frac{A}{2} \left[ 2 \pi \delta(f) + \frac{i e^{-2 \pi i f t_0 }  e^{-\pi^2 f^2  \sigma^2}}{\pi f}\right] .
        \label{eq:freq_glitch-model}
\end{equation}
The magnitude of $\tilde{g}$ falls like $1/f$ above the characteristic frequency $1/\sigma$.
This glitch model is depicted in Fig.~\ref{fig:glitch_model}.

\section{Background study: evidence for an underlying population of glitches}
Utilizing the glitch model in Eq.~\eqref{eq:glitch-model}, we analyze segments of LIGO data nearby GW231123's merger time. 
For each segment, we compute the Bayes factor comparing two hypotheses. The first is the hypothesis that there is only Gaussian noise in the data segment. The second, is the hypothesis that there is a microglitch, whose morphology is captured by Eqs.~\eqref{eq:glitch-model} and ~\eqref{eq:freq_glitch-model}, on top of the Gaussian noise. 
For the second hypothesis, we marginalize over the glitch parameters using uniform priors. The amplitude marginalization is performed analytically, the $\sigma$ marginalization is carried out numerically, and the time marginalization is carried out numerically using the fast Fourier transform trick described in \cite{Thrane:2019, Allen:2005fk}. 
These explicit marginalization steps are important to ensure that the sampler can find the microglitch and produce a well-converged evidence.
We refer the reader to appendix~\ref{sec:appendix-marginalization} for further details.

\begin{figure*}
    \centering
\includegraphics[width=0.49\textwidth]{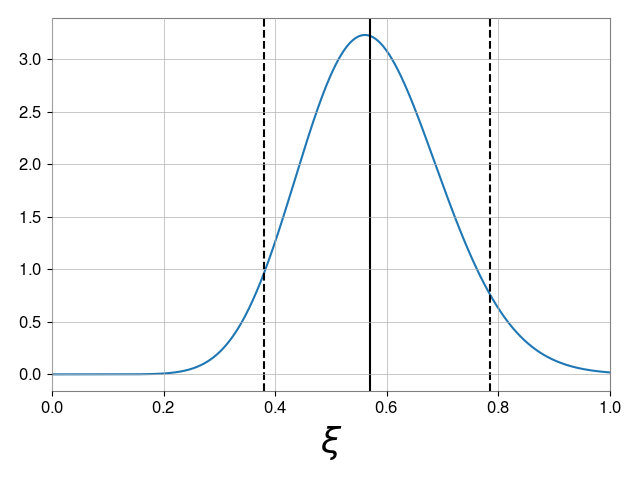}
\includegraphics[width=0.49\textwidth]{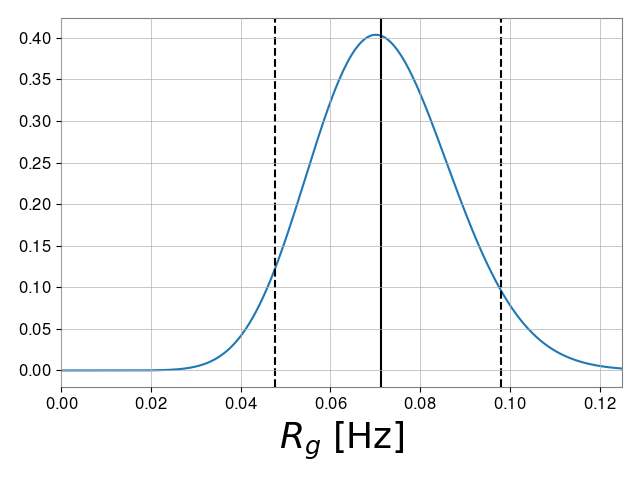}
\caption{The left-hand panel shows the posterior distribution for the fraction of segments near GW231123 that contain a microglitch. The right-hand panel is the same plot, but converted into a rate.}

    \label{fig:pop-study}
\end{figure*}

This list of $N$ Bayes factors can be used to estimate the fraction of data segments that contain a microglitch. We construct a Poisson mixture  model~\citep{Farr:2013yna} where the mixing faction $\xi$ represents the average fraction of segments with a microglitch.
The likelihood ${\cal L}$ for data $\vec{d}$ is given by:
\begin{equation}
    {\cal L}(\vec{d}|\xi, N) \propto \prod_i^N \left\{(1-\xi)+\xi\frac{{\cal L}(d_i|\mathrm{H}_{g})}{{\cal L}(d_i|\mathrm{H}_{n})}\right\}.
\end{equation}
Here, 
\begin{align}
    BF = \frac{{\cal L}(d_i|\mathrm{H}_{g})}{{\cal L}(d_i|\mathrm{H}_{n})},
\end{align}
is the Bayes factor for the hypothesis that there is a glitch in the $i^\text{th}$ data segment as opposed to just Gaussian noise.
Meanwhile, the microglitch-rate is computed from the duty cycle, and the duration~$(\Delta T)$ of each and the duration of each segment:
\begin{align}
    R_g = \frac{\xi}{\Delta T}
\end{align}
 The estimated posterior for $\xi$ is shown in Fig.~\ref{fig:pop-study}. We find $\xi = 0.57_{-0.19}^{+0.21}$ in the data around GW231123 which implies a microglitch rate of $R_g = \unit[0.07_{-0.02}^{+0.03}]{Hz}$ (both 90\% credibility), for a set of 250 data segments, each of which are $\Delta T = 8s$ long. We find that the  glitch rate depends only weakly on the choice of glitch parameter priors, with posteriors consistent between different values of the bounds on the uniform distributions used as population priors. We show in appendix~\ref{sec:appendix-sim} for how well this study can recover the underlying duty cycle from mock data comprising Gaussian noise realizations and glitch injections.


A potentially challenging aspect of explaining GW231123's spins with data quality issues that that both single-detector parameter estimation runs (i.e. with both LIGO Hanford and LIGO Livingston) show extremal spins. 
 Given that LIGO--Virgo--KAGRA detected 86 events in the most recent O4a observing run~\citep{LIGOScientific:2025slb}, and assuming that the microglitch rate is constant, the probability that at least one of these events is coincident with a microglitch in \textit{two} observatories simultaneously is 
\begin{align}
    1-(1-\xi^2)^{86} \sim 100\% .
\end{align}

We therefore conclude, that it is \textit{a prior} plausible that one event from the catalog has been affected by microglitches in both detectors. Thus we now move to investigating the potential impacts of microglitches on the astrophysical inferences of massive events like GW231123.

\section{Effects on astrophysical inferences}

\begin{figure*}[!ht]
   \centering \includegraphics[width=1.0\textwidth]{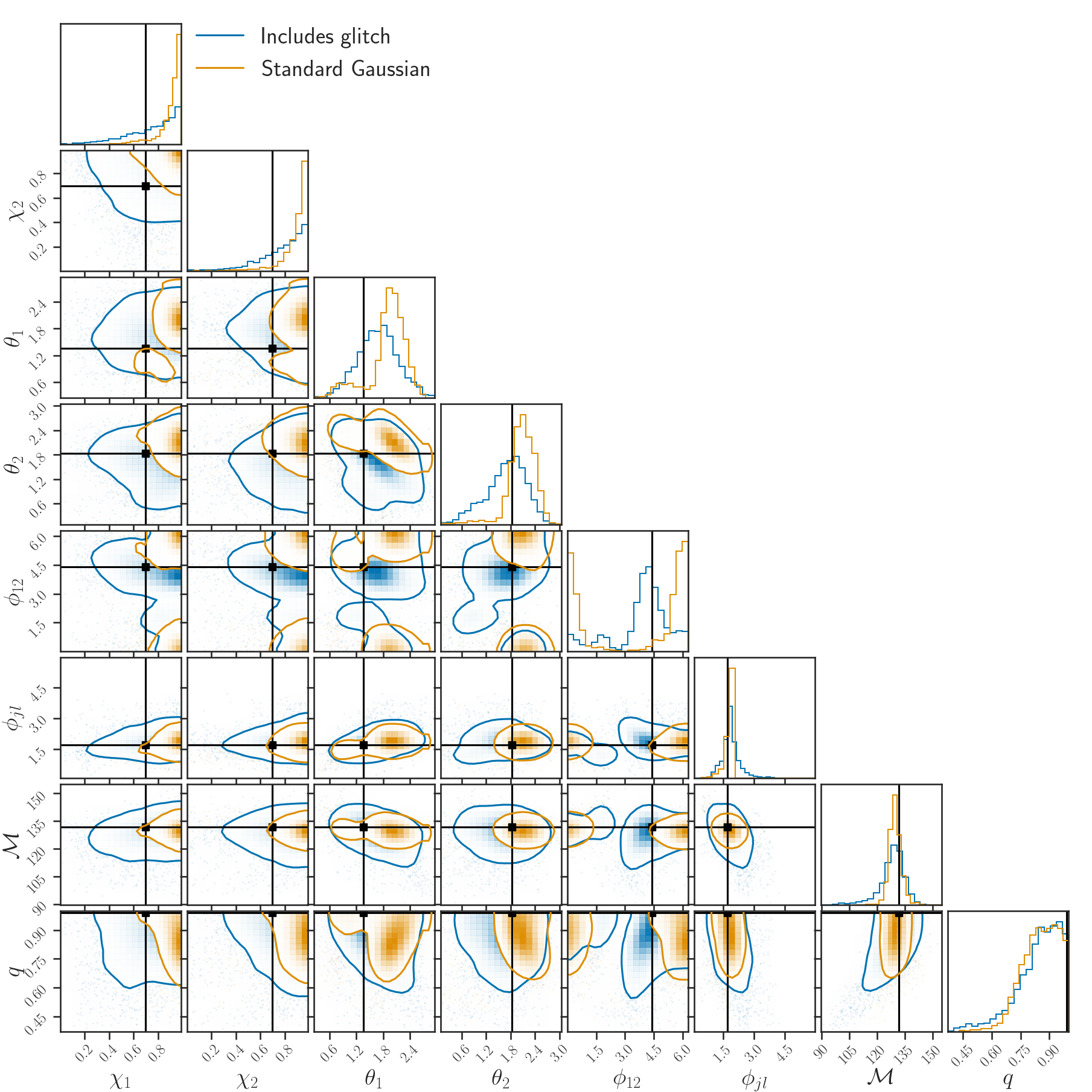}
   \caption{
   Posterior corner plot for a simulated GW231123-like event injected into Gaussian noise along with a microglitch.
   The true parameter values are marked in black. We show the distributions of chirp mass $\mathcal{M}$, mass ratio~$q$, component spin magnitudes $\chi_{1,2}$, and  tilts $\theta_{1,2}$, relative spin-orientation~$\phi_{12}$ and orientation between the total and orbital angular momentum $\phi_{JL}$.
   The orange distribution shows the posterior obtained using the standard Gaussian likelihood.
   Note that the true value is in many panels strongly excluded from the two-sigma credible intervals.
   In particular, the spin parameters peak strongly at $\chi=1$.
   The blue distribution shows the posterior obtained when we analyze the same data with our glitch model.
   The posterior broadens to include the true values.  We show all signal parameters except for luminosity distance which is marginalized and the fixed extrinsic parameters, namely sky position, coalescence phase, the polarization and inclination angles. For the standard Gaussian analysis we also marginalized over coalescence time.
   \label{fig:PE}}
\end{figure*}

To investigate how microglitches might bias astrophysical inference, we simulate a gravitational-wave signal with parameters taken to be the maximum a posteriori values reported for GW231123---\textit{except for the component spin magnitudes}.
We take the the dimensionless spin magnitudes to be $\chi_1=\chi_2=0.7$; consistent with what one might expect from hierarchical mergers in dense environments. 
The signal is generated with the \textsc{NRSur} approximant \citep{Varma:2019csw}.
The simulated signal, along with an error function glitch from~Eq~\ref{eq:freq_glitch-model}, is injected into a Gaussian noise realization of LIGO Hanford strain corresponding to the power spectral density measured for the GW231123 segment~\citep{ligo_scientific_collaboration_2025_16004263}. See the right panel of Figure~\ref{fig:glitch_model} for a visualization of how the microglitch can subtly alter (and hide behind) a short-duration CBC signal. Table~\ref{table:params} shows the injected gravitational-wave signal parameters as well as the injected glitch parameters. 

Using the same frequency band as the LVK analysis of GW231123~\citep{LIGOScientific:2025rsn}, the matched filter SNR of the glitch with data containing both the glitch and the CBC is 5.1, and the optimal SNR of the glitch is 1.12. In such a scenario, the presence of these glitches would not be confidently identified by pipelines such as \texttt{Gravity Spy}~\citep{Zevin:2016qwy, Glanzer:2022avx, Wu:2024tpr}; which in-particular only characterizes glitches above an omicron SNR of 8. However, we show that uncharacterized glitches of similar strength and duration, if present in the data, can significantly bias the PE of short-duration CBC signal. 
\begin{table}[h]
\centering
\begin{tabular}{cc}
\hline
\hline
Parameter & Value\\
\hline
$m_1$ & $152M_{\odot}$ \\
$m_2$ & $151M_{\odot}$ \\
$\chi_1$ & $0.7$ \\
$\chi_2$ & $0.7$ \\
$\theta_1$ & $1.4$ \\
$\theta_2$ & $1.8$ \\
$\phi_{12}$ & $4.4$\\
$\phi_{JL}$ & $1.7$\\
$\theta_{JN}$ & $1.4$\\
$d_L$ & $1173Mpc$\\
$\psi$ & $2.3$\\
$\alpha$ & $3.3$\\
$\delta$ & $0.3$\\
$\phi_c$ & $2.2$\\
$t_c$ & $1384782888.6s$\\
$\phi_c$ & $2.2$\\
$A$ & $5.9\times 10^{-22}$\\
$t_0$ & $1384782894.6s$\\
$\sigma$ & $10^{-4}$\\
\hline
\end{tabular}
\caption{Parameter values of the CBC signal, namely component masses~$(m_{1,2})$, dimensionless spin magnitudes~$(\chi_{1,2})$, spin tilt angles~$(\theta_{1,2})$, relative spin orientaion~$(\phi_{12})$, orientation between orbital and total angular momentum~$(\phi_{JL})$, inclination angle~$(\theta_{JN})$, luminosity distance~$(d_L)$, polarization angle~$(\psi)$, right ascension~$(\alpha)$, declination angle~$(\delta)$, phase at coalescence~$(\phi_c)$ and time at coalescence in geocentric co-ordinates~$(t_c)$, and the the glitch~$(A,t_0, \sigma)$. Among the dimensionful quantities, all angles, orientations, and sky positions are expressed in radians. }
\label{table:params}
\end{table}


We perform parameter estimation on this data using two different likelihood models. The first is with the standard Gaussian Whittle likelihood assuming only the presence of a gravitational-wave signal. The second likelihood assumes that there is both a binary black hole signal and a microglitch with the specific form of Eq.~\ref{eq:freq_glitch-model}(but otherwise the noise is Gaussian). The dimensionality of the parameter space when the three glitch parameters are included with the fifteen binary parameters, makes the  computationally difficult.
In order to obtain an illustrative result that can demonstrate the ability of microglitches to bias parameter estimation, we therefore fixed some of the extrinsic binary parameters: right ascension, declination, phase of coalescence, inclination, and polarization angles. The luminosity distance and time of coalescence ---along with the eight intrinsic parameters of the gravitational-wave signal were allowed to vary. A more complete analysis is in progress, though, we believe the results will be qualitatively similar to the ones shown here.
We explicitly marginalize over glitch amplitude, glitch time, and luminosity distance using methods similar to the glitch-only scenario~(see Appendix~\ref{sec:appendix-marginalization}).
In order to control computational costs, we fix $\sigma$ to its true value.
We anticipate that the result is similar to what we would obtain if we marginalized over $\sigma$ because the likelihood is does not resolve this parameter well when we analyze the simpler problem with just a glitch but no gravitational-wave signal (see appendix~\ref{sec:appendix-glitchpe}). 
The resulting posteriors are shown in Fig.~\ref{fig:PE}.

We find that the usual likelihood leads to biased recovery of the injected parameters. 
In particular, we find the posterior support for the component spin magnitude rails at $\chi=1$, with no support for the injected value of $0.7$, the $90\%$ credible intervals of $\chi_{1}$ and $\chi_2$ being $(0.78,0.99)$ and $(0.76,0.99)$ respectively. 
Biases also appear in the component split-tilt angles~$(\theta_{1,2})$ and relative spin-orientation $\phi_{12}$. 
On the other hand, the likelihood that includes the glitch produces a posterior distribution consistent with the true parameter values.
There is much higher posterior support at the injected value for all the binary parameters, including the component spin magnitudes $\chi_{1,2}$, whose $90\%$ credible intervals of $(0.47,0.99)$ and $(0.52, 0.99)$ for the primary and secondary respectively, now enclose the true values. Higher posterior support at the true values is also observed for the relative spin orientation. The glitch likelihood is favored over the Gaussian noise likelihood with a Bayes factor of $370$. 

We conclude the extreme spins inferred for GW231123 by \cite{LIGOScientific:2025rsn} could be a result of unmodeled microglitches.
The black holes in this system might have modest spins of $\chi\approx0.7$ if the inference calculation has been affected by plausible nonstationary noise.
Work is ongoing to reanalyze GW231123 with our microglitch model to provide updated parameter estimation results.
However, the calculation is computationally difficult, and so we are sharing these preliminary results in the meantime.

\section{Conclusion and Future prospects}
In this paper, we investigate the possibility that weak but frequent nonstationary noise events, which we call \textit{microglitches}, might bias astrophysical inferences about GW231123. 
Building on work from \cite{shun_memory}, we illustrate the existence of a population of microglitches with a step-function morphology in the data segments surrounding GW231123. We show that microglitches are sufficiently common to impact data in both LIGO observatories, and in particular we estimate a rate around GW231123 of $R_g = \unit[0.07_{-0.02}^{+0.03}]{Hz}$. A couple of caveats are in order here; firstly the general glitch rate in O4a was fairly high~\citep{LIGO:2024kkz}. Louder glitches that would be be gated or removed in an on-source analysis might be contaminating some of our microglitch rate measurement. Second, some microglitches are possibly spurious, arising from an frequency correlations induced by finite FFT windows~\citep{Talbot:2021igi, Talbot:2025vth} or a biased power spectral density estimation. Several techniques have been suggested in the literature for accounting for the later~\citep{Banagiri:2019lon, Talbot:2020auc, Biscoveanu:2020kat} and can be integrated into both our background estimation steps and microglitch-included likelihood.

 We also show signal coincident with glitches can yield biased astrophysical inference if special care is not taken to model non-stationary noise.
In particular, we show that it is possible to obtain posteriors that favor maximum component spins, even if the true signal contains black holes with more modest spins of $\chi\approx0.7$. 
We show that one can model non-Gaussian noise in order to eliminate this bias.

Unfortunately, the more complicated noise model is computationally expensive, and so our analysis here is only indicative.
Analysis of GW231123 (and other massive events through GWTC-4) with our microglitch model is underway. 
Looking further into the future, we see a number of ways to improve our model, for example, by using hierarchical inference to determine the prior distributions of glitch parameters.

\section{Acknowledgements}
We thank Simona Miller for a very constructive internal review of our manuscript. We are grateful to Sylvia Biscoveanu, Colm Talbot, and Charlie Hoy for insightful discussions. AR is supported by the National Science Foundation~(NSF) award PHY-2207945 and acknowledges the support of the NSF-Simons AI-Institute for the Sky (SkAI) via grants NSF AST-2421845 and Simons Foundation MPS-AI-00010513. 
 S.B. and E.T. are supported by the Australian Research Council CE230100016, LE210100002, DP230103088. 
We are grateful for the computational resources provided by the LIGO laboratory and supported by National Science Foundation Grants PHY-0757058 and PHY-0823459. This
material is based upon work supported by NSF’s LIGO Laboratory, which is a major facility fully funded by the National Science Foundation.
\appendix
\section{Explicitly marginalized likelihood}
\label{sec:appendix-marginalization}
The glitch-inclusive likelihood can be explicitly marginalized over several parameters, which is necessary either for tractable and well-convergent PE or for computing the Bayes-factors required by our background study. In the absence of CBC signals, the glitch-inclusive log-likelihood-ratio takes the following functional form:
\begin{equation}
    \mathcal{L}^g = A\left<d|\bar{g}(\sigma,t_g)\right> - \frac{1}{2}A^2 \left<\bar{g}(\sigma,t_g)|\bar{g}(\sigma,t_g)\right>\label{eq:glitch-likelihood},
\end{equation}
where $\bar{g}=\frac{g}{A}$ is the normalized glitch frequency series. Imposing a uniform prior on the glitch amplitude bounded by $(-A_0,A_0)$, we can analytically marginalize Eq.~\eqref{eq:glitch-likelihood} to get:
\begin{eqnarray}
     \mathcal{L}^g_A(\left<d|\bar{g}\right>, \left<\bar{g}|\bar{g}\right>) &=& \frac{2A_0}{A_{max}}\left\{ \frac{1}{\sqrt{\left<\bar{g}|\bar{g}\right>}} 
e^{\left( \frac{0.5 \, \left<d|\bar{g}\right>^2}{\left<\bar{g}|\bar{g}\right>} \right)} 
\left( 
-\sqrt{\frac{\pi}{2}} \, \mathrm{Erf} \left( 
\frac{ -\frac{1}{\sqrt{2}} \left<d|\bar{g}\right> - \frac{1}{\sqrt{2}} \, A_{\mathrm{max}} \, \left<\bar{g}|\bar{g}\right> / A_0 }{ \sqrt{ \left<\bar{g}|\bar{g}\right> } } 
\right) \right.\right.\nonumber \\
&+&\left.\left. \sqrt{\frac{\pi}{2}} \, \mathrm{Erf} \left( 
\frac{ -\frac{1}{\sqrt{2}} \left<d|\bar{g}\right> + \frac{1}{\sqrt{2}} \, A_{\mathrm{max}} \, \left<\bar{g}|\bar{g}\right> / A_0 }{ \sqrt{ \left<\bar{g}|\bar{g}\right> } } 
\right) 
\right)\right\}
\end{eqnarray}
where the subscript $_A$ represents marginalization over $A$. We can further marginalize over time using the fast Fourier transform:
\begin{equation}
\mathcal{L}^g_{A,t_g} = \log\left(\pi(t_k) \sum_k \exp\{\mathcal{L}_A(\left<d|\bar{g}(t_k)\right>, \left<\bar{g}|\bar{g}\right>)\}\Delta t\right),
\end{equation}
where,
\begin{equation}
        \left<d|\bar{g}(t_k)\right> = 4 \Delta f \mathcal{R} \mathrm{fft}_k \left(\frac{d_j \bar{g}_j(t_0)}{P_j}\right),
\end{equation}
and $t_0$ is some reference time. From this likelihood, one can obtain the Bayes factor in favour of the glitch hypothesis over that of only Gaussian noise by numerically integrating out $\sigma$ on a grid.

Similarly, in the presence of CBC signals, the log-likelihood-ratio can be expressed as an explicit function of glitch amplitude, time, and signal luminosity distance~$(D_L)$:
\begin{equation}
    \mathcal{L}^{g,h} = \frac{D_0}{D_L} \left<d|\bar{h}(\vec{\theta})\right> +  A\left<d|\bar{g}(\sigma,t_g)\right> -\frac{1}{2}\left\{A^2\left<\bar{g}(\sigma,t_g)|\bar{g}(\sigma,t_g)\right>+\left(\frac{D_0}{D_L}\right)^2\left<\bar{h}(\vec{\theta})|\bar{h}(\vec{\theta})\right> + 2 A \frac{D_0}{D_L} \left<\bar{h}(\vec{\theta})|\bar{g}(\sigma,t_g)\right>\right\}, \label{eq:glitch-cbc-likelihood}
\end{equation}
where $D_0$ is some reference distance, and $\bar{h}$ is the CBC signal model evaluated at $D_0$ as a function of other signal parameters $\vec{\theta}$. Similar to the glitch-only likelihood, we can analytically marginalize Eq.~\eqref{eq:glitch-cbc-likelihood} over $A$ to obtain:
\begin{align}
\mathcal{L}^{g,h}_A(\left<d|\bar{g}\right>, \left<\bar{h}|\bar{g}\right>, \left<\bar{h}|\bar{h}\right>, \left<\bar{g}|\bar{g}\right>, D_L) &= \nonumber \\
& ~
\Bigg(
  \tfrac{1}{2}\,\left< d \mid \bar{g} \right>^{2}
  - \left< d \mid \bar{g} \right>\, \left< \bar{g} \mid \bar{h} \right>\,
    \left(\tfrac{D_{0}}{D_{L}}\right) + \left(\tfrac{D_{0}}{D_{L}}\right)\!\Big(
      \left< d \mid \bar{h} \right>\, \left< \bar{g} \mid \bar{g} \right>
      + \tfrac{1}{2}\, \left< \bar{g} \mid \bar{h} \right>^{2}
        \left(\tfrac{D_{0}}{D_{L}}\right) \nonumber \\
& \quad\qquad
      - \tfrac{1}{2}\, \left< \bar{g} \mid \bar{g} \right>\,
        \left< \bar{h} \mid \bar{h} \right>\,
        \left(\tfrac{D_{0}}{D_{L}}\right)
    \Big)
\Bigg)
/ \left< \bar{g} \mid \bar{g} \right>
\nonumber \\
& \quad
+ \log \Bigg\{
  \frac{1}{\sqrt{\left< \bar{g} \mid \bar{g} \right>}}
  \Bigg[
      -\tfrac{2}{\sqrt{\pi}} \, \operatorname{Erf}\!\left(
        \frac{-\left< d \mid \bar{g} \right>
              - A_{\max}\, \left< \bar{g} \mid \bar{g} \right>
              + \left< \bar{g} \mid \bar{h} \right>\,
                \left(\tfrac{D_{0}}{D_{L}}\right)}
             {\sqrt{2 \left< \bar{g} \mid \bar{g} \right>}}
      \right) \nonumber \\
& \qquad\;\;\;
    + \tfrac{2}{\sqrt{\pi}} \, \operatorname{Erf}\!\left(
        \frac{-\left< d \mid \bar{g} \right>
              + A_{\max}\, \left< \bar{g} \mid \bar{g} \right>
              + \left< \bar{g} \mid \bar{h} \right>\,
                \left(\tfrac{D_{0}}{D_{L}}\right)}
             {\sqrt{2 \left< \bar{g} \mid \bar{g} \right>}}
      \right)
  \Bigg]
\Bigg\},
\end{align}
which can again be marginalized over glitch-time using the fast Fourier transform as follows:
\begin{equation}
    \mathcal{L}^{g,h}_{A,t_g} = \log\left(\pi(t_k) \sum_k \exp\{\mathcal{L}^{g,h}_A(\left<d|\bar{g}(t_k)\right>, \left<\bar{h}|\bar{g}(t_k)\right>, \left<\bar{h}|\bar{h}\right>, \left<\bar{g}|\bar{g}\right>, D_L)\}\Delta t\right),
\end{equation}
where $\left<d|\bar{g}(t_k)\right>$ and $t_0$ are the same as before, and
\begin{equation}
        \left<h|\bar{g}(t_k)\right> = 4 \Delta f \mathcal{R} \mathrm{fft}_k \left(\frac{h_j \bar{g}_j(t_0)}{P_j}\right).
\end{equation}
Finally, we marginalize over $D_L$ numerically on a grid. Note that we could have, in principle, also marginalized over CBC coalescence time by employing a two-dimensional fast Fourier transform. However, for typical values of the sampling frequency used in PE, such computations are intractable. Nevertheless, the marginalized log-likelihood ratio can be tractably sampled using suitable priors on the remaining CBC parameters even without marginalization over CBC coalescence time.

\section{Reconstructing Glitch Parameters}
\label{sec:appendix-glitchpe}
\begin{figure*}
    \centering
    \includegraphics[width=0.95\linewidth]{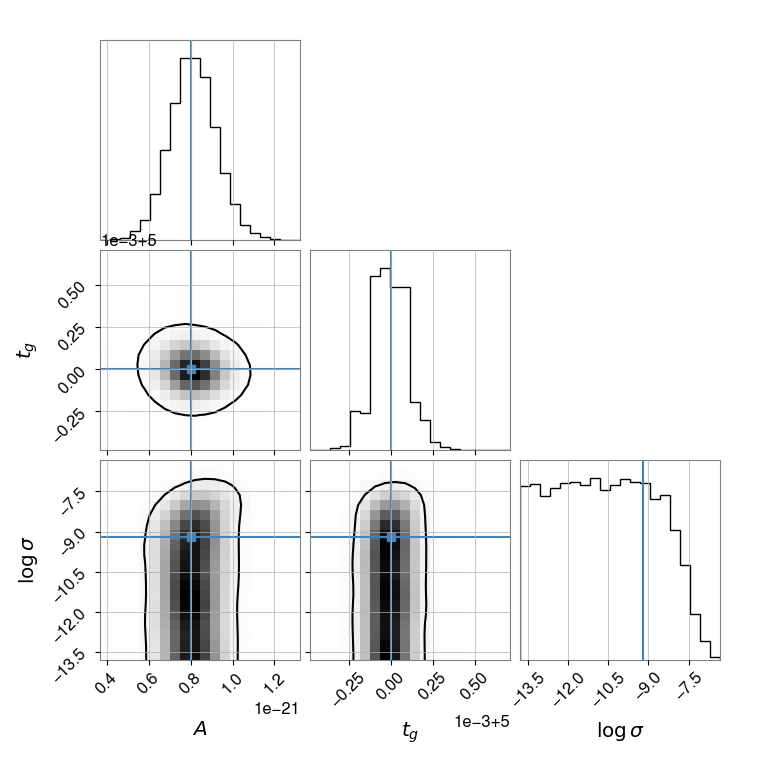}
    \caption{Reconstructed posterior samples of the marginalized parameters, for an injected glitch, with true values marked in blue.}
    \label{fig:glitch_rec}
\end{figure*}

Once posterior samples of the unmarginalized CBC parameters are obtained, they can be used to reconstruct the marginalized glitch parameters using methods similar to the one described in \cite{Thrane:2019}. This is demonstrated for an injected glitch in Gaussian noise in figure~\ref{fig:glitch_rec}. Note that the inference is marginally informative on $\sigma$.

\section{Simulation recovery for Background Study}
\label{sec:appendix-sim}
\begin{figure*}
    \centering
\includegraphics[width=0.49\textwidth]{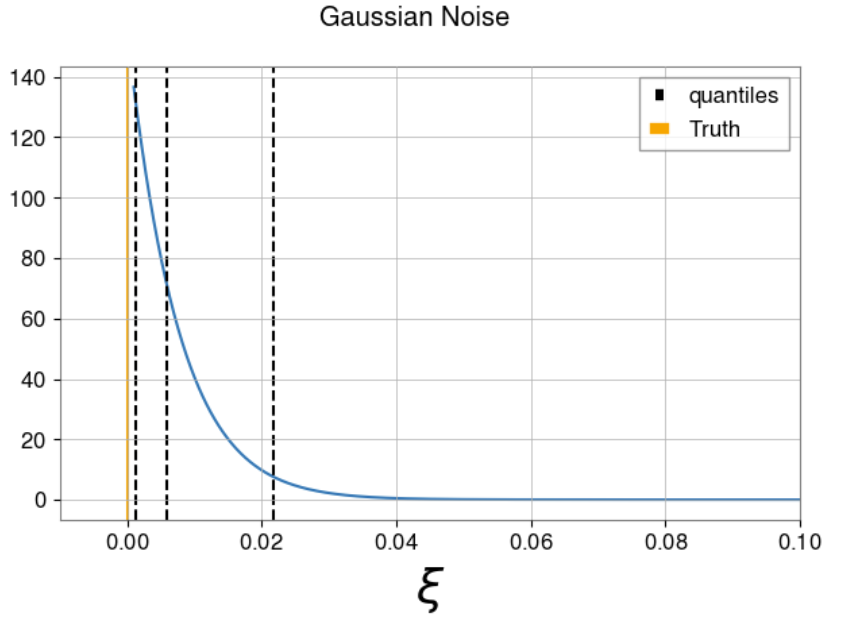}
\includegraphics[width=0.49\textwidth]{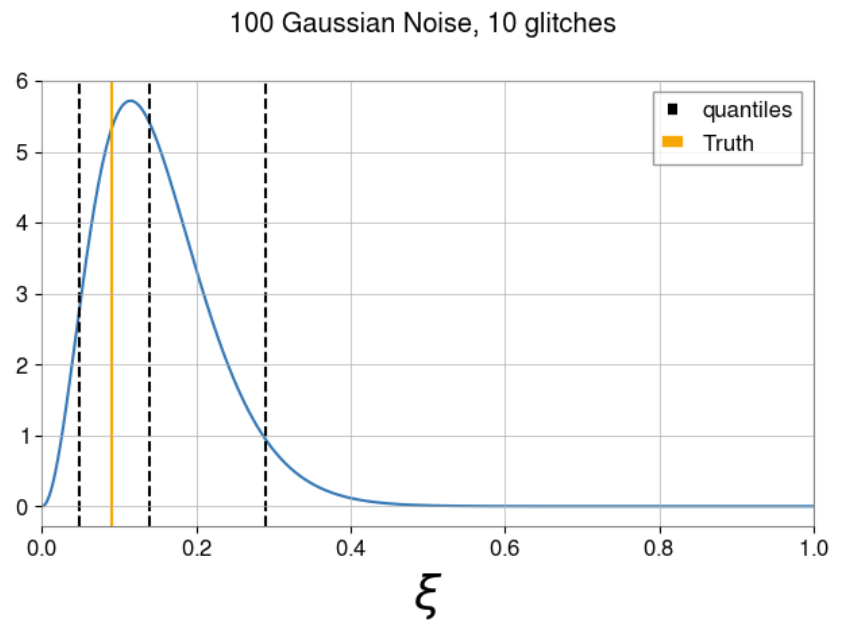}
\caption{The left panel shows the inferred duty cycle from only Gaussian noise realizations and the right-hand panel shows the same but from the case with 10 glitch injections.}

    \label{fig:pop-study-sim}
\end{figure*}

Here we re-do our Background study on two sets of mock data, one that comprises of 100 Gaussian noise realizations and the other one has 10 additional glitch injections. As shown in Figure~\ref{fig:pop-study-sim}, our background study recovers the correct value of the glitch duty cycle in both cases.
\bibliography{lvk_refs, references}

\onecolumngrid
\include{supplementary}

\end{document}